# Finite Element Investigation of the Deterioration of Doweled Rigid Pavements

**Ziad Ghauch**

Undergraduate student, Department of Civil Engineering, Lebanese American University, Byblos, e-mail: zdghaouche@gmail.com

## Abstract

This study investigates the failure of concrete around dowel bars in jointed rigid pavements, and the resulting effect on the pavement performance. In fact, under repetitive vehicle loading, concrete in contact with the dowel bar deteriorates, particularly at the joint face. The degradation of concrete around the dowel negatively affects the latter's performance in terms of transferring wheel loads through vertical shear action. In this context, a nonlinear 3D Finite Element (FE) analysis was performed using the commercial FE code ABAQUS (v-6.11). The 3D FE model includes friction interfaces, infinite boundary elements, and 3D beam model for dowel bars. The obtained numerical results were validated with classical analytical solutions of shear and moment along the dowel. A concrete damaged plasticity model was used for the concrete slab to model the degradation of concrete matrix around the dowels under incremental loading. Results obtained show, among other things, that the degradation of concrete around dowel bars was found to initiate at the face of the joint and propagate towards the interior of the dowel. Also, the central dowels under the wheel load lost a significant portion of their load-transfer capacity as the concrete matrix around them deteriorated, while dowels farther away from the wheel load became more engaged in load transfer. Finally, it was confirmed that the overall vertical load transferred by all the dowels across the joint decreases as the concrete matrix deteriorates.

**Keywords:** Jointed concrete pavements; dowel bars; nonlinear Finite Element analysis.

# Introduction

**Background**

The use of joints in rigid pavements allows thermal expansion and contraction of the concrete slabs at the expense of breaking the continuity of the pavement structure. Inserting dowel bars between concrete slabs has proved beneficial in reducing distresses at pavement discontinuity such as corner cracking and joint faulting (Yu et al. 1998, Hoerner et al. 2000). The primary function of dowel bars is to transfer vertical load from the loaded to the unloaded PCC slab through shear action, hence reducing maximum deflections and critical stresses.

The study of the performance of concrete pavements and dowel bars through the finite element method is not new. Several attempts to model load transfer between adjacent slabs have been proposed in the past, as shown in table [1]. Each method has its own advantages and inconveniences. An early attempt to model load transfer was by connecting the faces of the PCC joint with linear spring elements with stiffness assigned in the vertical direction (Huang and Wang 1973). Guo et al. (1995) presented a component dowel bar model that consists of a shear bending beam connecting two bending beams in the concrete matrix. Dowel-concrete interaction was modeled via the stiffness of springs connecting the dowel bar to the surrounding concrete. The ILLI-SLAB model presented by Tabatabaie et al. (1979) considers the dowel bar as a beam element and the support provided by the PCC matrix as a single spring that acts at the joint face and whose stiffness is the dowel-concrete interaction (DCI) value. In a nonlinear 3D finite element model, Channakeshava et al. (1993) modeled the dowel bar with 3D beam elements, while the dowel-concrete interaction was modeled with spring elements connecting the dowel nodes to the concrete solid elements.

Under repetitive wheel load passages, the concrete in contact with the dowel bar deteriorates, particularly at the joint face. The high levels of stress in the concrete around the dowels, coupled with relatively limited bearing strength of concrete, causes the deterioration of the concrete matrix around the dowels due to fatigue loading (Channakeshava et al. 1993), which negatively affects the performance of the pavement structure. The formation of voids following the degradation of the concrete around the dowels depends, among other things, on the pavement structural properties, the number of load applications, and the bearing strength of the supporting concrete.

Considerable research is found on modeling load transfer devices across PCC joints (see table [1]). However, little effort has been spent on describing the deterioration of concrete around the dowels resulting from repetitive wheel loading. An early attempt to model the degradation of concrete around the dowels was presented by Larralde (1984). The modulus of Dowel-Concrete Interaction (DCI) can be calculated using the following equation:

$$DCI = \frac{K^{0.75} \phi^{2.5}}{0.041 \phi^{2.5} + 0.0004 K^{0.25} z} \quad [1]$$

Where ø is the diameter of dowel bar, K is the modulus of dowel support, and z is the joint width. In order to model the deterioration of the concrete around the dowels and the formation of voids, a reduction factor is applied to the DCI value. The reduction factor (RF) can be calculated as follows:

$$RF = 0.268 - 0.0457 \, Log \, (N) + 1.123 \, f_{rb} \quad [2]$$

Where N is the number of load repetitions, and $f_{rb}$ is the relative load acting on the dowel, function of the concentrated force acting on the dowel, the thickness of the slab, the compressive strength of concrete, the diameter and the embedment length of dowel, and the joint width.

Several researchers have observed that the deterioration of concrete around the dowel under repetitive loading is mostly limited to the face of the joint. Channakeshava et al. (1993) noted that "*most of the shear is transferred through the end interface spring, and the interior springs remain linear in response.*" This leads to the idea that the degradation of the concrete material surrounding the dowel bar is not uniform along the embedded section of the dowel. A new degradation model for the concrete around dowel bars, that takes into account local damage effects, needs to be established.

**Objective and Scope**

This study represents a numerical investigation of the deterioration of concrete around dowel bars in typical rigid pavement structures, and the consequent effect on the pavement performance. The first section describes the failure of concrete matrix around dowel bars and investigates whether this failure of concrete matrix occurs through local or general degradation. The second part of this study examines the impact of concrete matrix degradation on the modulus of dowel support K along dowel bars. Finally, attempts are made to quantify the reduction in load transfer capacity of each dowel bar as the surrounding concrete deteriorates, and to underline the mechanisms of dowel group action with deteriorating concrete matrix.

*Table 1 – Summary of major attempts to model load transfer mechanisms in doweled joints*

| Model for doweled joint | Reference | Low computational cost | Include model for deterioration of dowel support | Inconvenient(s) |
|---|---|---|---|---|
| Linear elastic spring with vertical stiffness connecting concrete slabs | Huang and Wang (1973) | ✓ | ✓ | Elastic spring provides no bending resistance |
| Timoshenko beam element for dowel bar connecting adjacent slabs modeled with Kirchhoff plate elements | Tabatabaie (1978) | | | Embedded portion of dowel bar is not considered |
| Timoshenko beam element for dowel connected to concrete slab through elastic springs | Guo et al. (1995) | | ✓ | Dowel behavior greatly dependent on elastic spring constant, K, and the K-value shows great variability |
| Dowel bar modeled as beam element embedded in continuum elements of concrete slab | Davids et al. (2003) | ✓* | ✓ | |

*Embedded beam element formulation for the dowel allows significant savings in computational cost

## Numerical Modeling of Pavement Structure

In this first section, a detailed description of the 3D FE model of the pavement structure is presented. Particularly, the geometry, loads and boundary conditions of the pavement model are discussed. Then, non-linear material properties for the concrete slab are reviewed; and finally, the developed FE model is validated with classical solutions presented by Timoshenko (1925) and Friberg (1938).

**Model Geometry, BC's, and Loads**

A 3D Finite Element (FE) model of the rigid pavement structure was developed using ABAQUS 6.11 (ABAQUS 2011). The rigid pavement sections consists of a 220 mm Portland Cement Concrete (PCC) slab, overlaying a 300 mm aggregate base, on top of an elastic solid subgrade soil. An illustration of the FE mesh of the pavement is shown in figure [1]. The modeled section of the rigid pavement was 13,032 mm in the longitudinal direction, and 5,086 mm in the transverse direction. A transverse joint width of 10 mm was assumed in order to allow expansion and contraction of the PCC slabs. At this relatively high joint width, load transfer is only achieved through dowel bar action, and the effect of aggregate interlock on amount of load transfer can be safely neglected. Due to the symmetry of the model along the longitudinal central axis, only half of the pavement section was modeled, and symmetric BC's were inserted along the vertical plane of symmetry in the traffic direction. Boundaries of the PCC slab in the traffic direction were not constrained. A finer mesh was implemented around the loading area, and in the zones of high stress/strain gradients, while relatively larger elements were used farther away. The mesh was refined several times until the point where further mesh refinement resulted in little or no change in solution. The length of the smallest element used in the FE model was of the order of 7.5 mm. The interface between the PCC slab and the aggregate base layer was modeled using frictional contact.

The behavior of the aggregate base layer was modeled using the Mohr-coulomb plasticity model, as shown in figure [1]. The subgrade soil was modeled as a linear elastic material, with Young's modulus of 60 MPa, and Poisson's ratio of 0.4. The dowel bar was modeled as an elastic-plastic material, with a Young's modulus of 200 GPa, a Poisson's ratio of 0.3, and a yield strength of 275 MPa.

In the 3D FE pavement model, infinite elements were inserted along the vertical and horizontal boundaries of the subgrade soil, as shown in figure [2]. The use of infinite boundary elements in problems involving infinite domains has proved beneficial in reducing the computational time of FE analyses (Bettess and Zienkiewicz 1977). This is due to the fact that a large number of far-field finite elements used to extend the boundaries to adequate distances so as to achieve approximately zero displacements at the boundaries can be replaced with infinite elements. In addition, the location of boundary infinite elements is crucial. They must be located as close as possible to the wheel load in order to minimize the size of the FE analyses, but not too close to alter the results in the zone of interest (ARA 2004). The location of infinite elements at the bottom of the subgrade soil was selected at a depth where the vertical stress became less than 1% of the applied pressure.

A 40 KN load ($F_{40}$), with circular tire imprint of 300 mm diameter, was applied on the edge of one PCC slab. Gradually, the load (F) was increased until all of the concrete surrounding the dowel bars under the applied load fully deteriorated. An automatic incrementation scheme was adopted, with a maximum increment size of 0.0001 seconds. Although it may have been possible to take a larger maximum increment size with many more iterations for each increment, the solution with smaller increments was found more efficient. Due to the large number of nonlinearities involved in the system, automatic stabilization was defined for the purpose of improving the rate of convergence of the solution. The Quasi-Newton incremental/iterative technique was used as it provides substantial savings in computational cost, as opposed to the Newton-Raphson technique, particularly for small-displacement analysis with only local plasticity.

Dowel bars connecting adjacent PCC slabs were modeled using 3D beam elements. The dowel is 32 mm in diameter, and is embedded 215 mm in the PCC slab. The dowel nodes were connected to the surrounding concrete using nonlinear spring elements. In compression, a relatively high stiffness was assigned for the springs in order to simulate perfect bond between the dowels and the surrounding concrete, while in tension, the springs were allowed to deform freely (zero stiffness). Such a bilinear behavior was found to best replicate field conditions.

**Material Properties**

The mechanical behavior of concrete is modeled using the concrete damaged plasticity model (Lubliner et al. 1989, Lee and Fenves 1998). An isotropic damaged elasticity coupled with isotropic compressive and tensile plasticity is used in this continuum damage model for concrete. The concrete damaged plasticity model assumes different yield strengths in tension and compression. In compression, initial hardening is followed by softening behavior whereas in tension, only softening behavior occurs.

The post-failure tensile behavior was specified in terms of a stress-displacement curve (Lee and Fenves 1998), while tensile damage was specified in terms of cracking displacement. Values for the parameters of the concrete damaged plasticity model were obtained from Bhattacharjee et al. (1993) & Ghrib and Tinawi (1995) and are shown in figure [1]. While the failure mechanisms of concrete under low confining pressures are cracking in tension and crushing in compression, the stiffness degradation $d_c$ caused by compressive crushing was neglected.

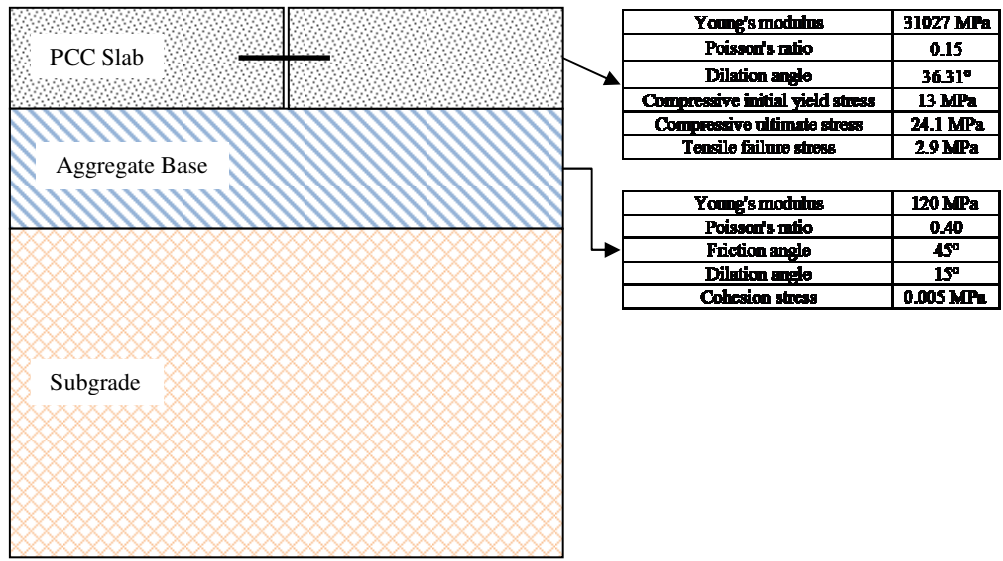

*Figure 1 – Pavement profile and material properties for PCC slab and aggregate base layer*

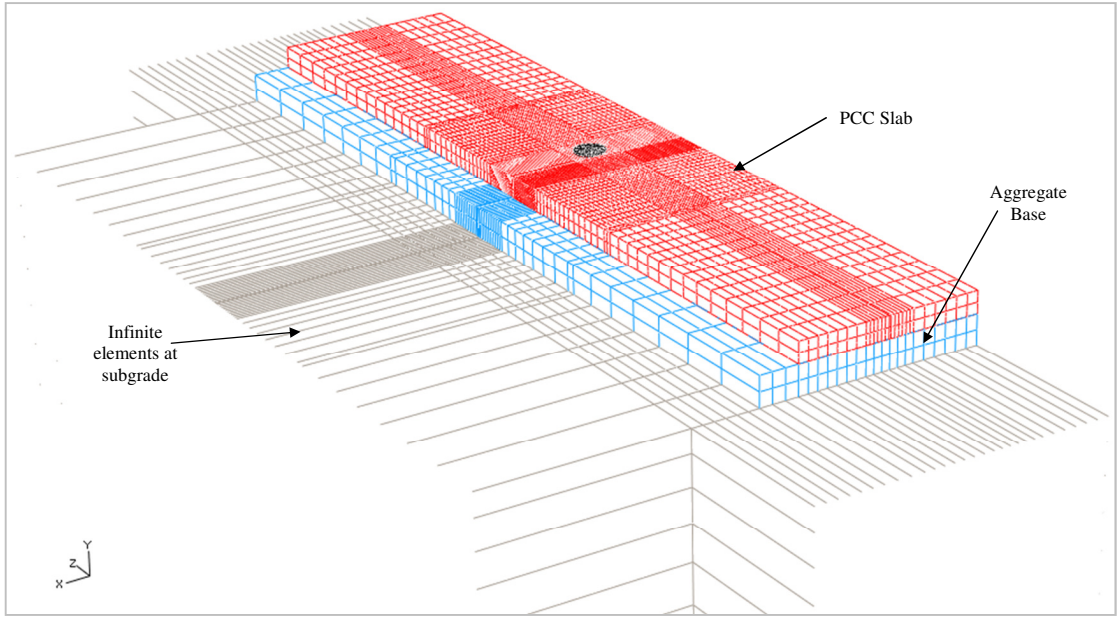

*Figure 2 – 3D FE mesh of rigid pavement section*

**FE Model Validation with Analytical Solutions**

The 3D FE model was validated with analytical solutions for shear force and moment along the dowel bar presented by Timoshenko and Lessels (1925) and by Friberg (1938). The general equation for the deflection of a dowel structure extending infinitely into an elastic mass can be written as follows (Timoshenko 1925):

$$y(x) = \frac{e^{-\beta x}}{2\beta^3 E_s I}[P \cos\beta x - \beta M_0 (\cos\beta x - \sin\beta x)] \qquad [3]$$

Where β is the relative stiffness of the dowel structure and is defined as follows:

$$\beta = \sqrt[4]{\frac{K\emptyset}{4 E_s I}} \qquad [4]$$

In equation [4], $E_s$ and I are the modulus of elasticity and moment of inertia of the dowel bar, respectively, ø is the diameter of dowel bar, and K is the modulus of dowel-support, representing the pressure intensity (MPa) required to induce a 1mm settlement, and P is the downward shear acting on the dowel, and transferred from one PCC slab to the next.

In order to validate the obtained numerical solution with the classical, a proper value of β should be used in the analytical expressions. The value of β was calculated as follows: once the deflection of the dowel at the face of the joint ($y_o$) and the maximum shear acting on the dowel (P) are obtained from numerical results, the following equation can be solved for β:

$$y_o = \frac{P}{4\beta^3 E_s I}[2 + \beta z] \qquad [5]$$

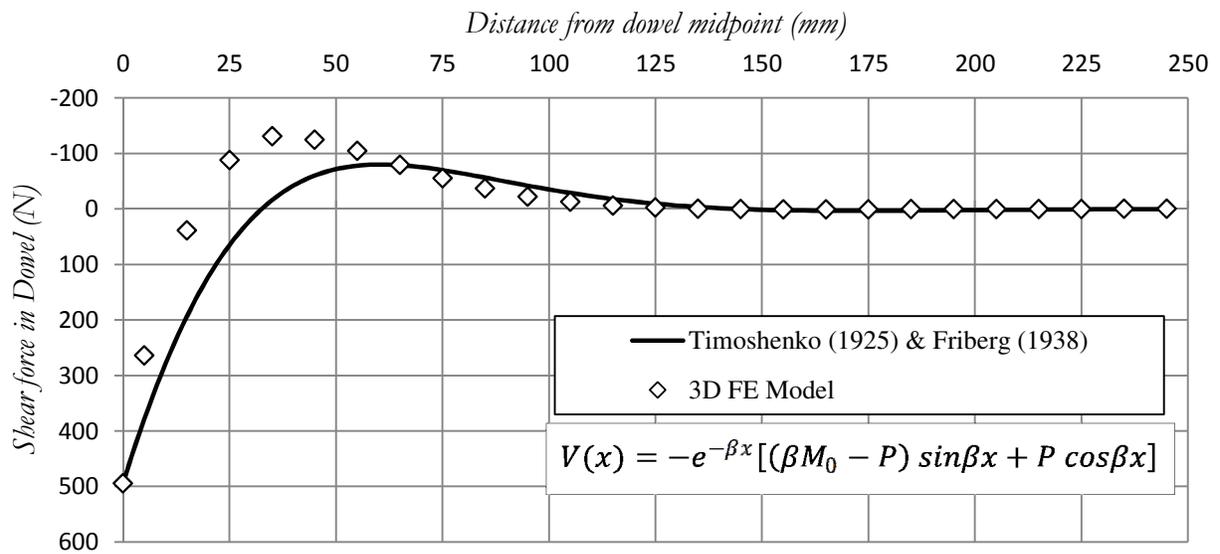

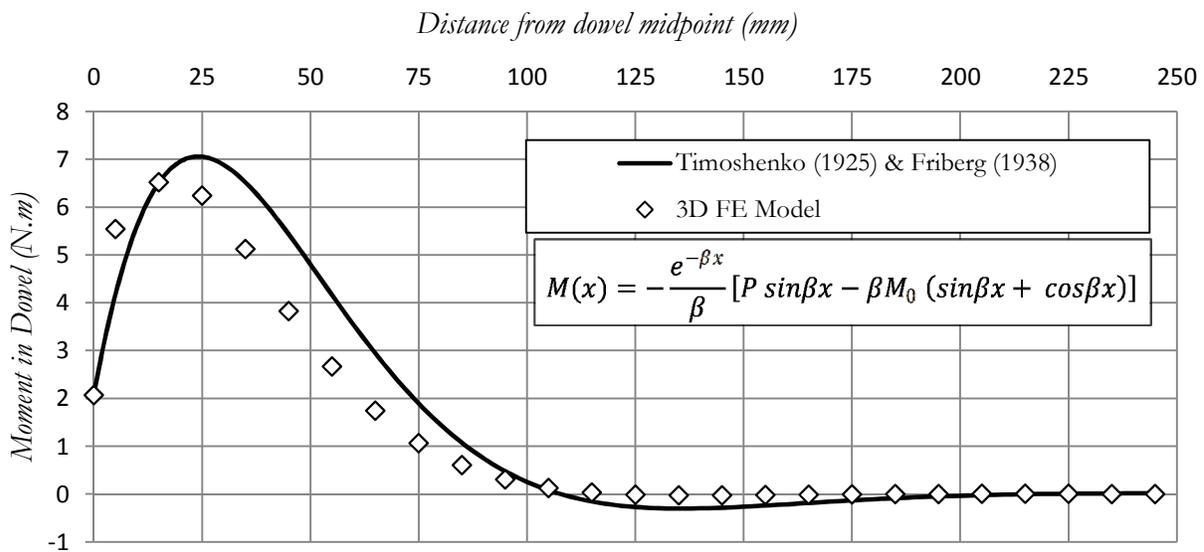

*Figure 3 – Validation of 3D FE model with classical solutions at the dowel directly under the applied load for (a) shear force, and (b) moment along the embedded section of the dowel for β = 0.028 $mm^{-1}$*

The modulus of dowel support K can be calculated from equation [4] once the relative stiffness of the dowel ($\beta$) is determined. Knowing $\beta$, the shear V(x) and moment M(x) in the dowel bar obtained from equation [3] can be solved, as shown in figures [3a] and [3b] respectively. Friberg (1938) assumed a point of counter-flexure exists at the dowel midpoint, hence expressing the moment $M_o$ at the face of the joint as a function of P and the joint width z as follows: $M_0=Pz/2$.

Figure [3] shows a comparison of the analytical solution of Timoshenko (1925) and Friberg (1938) with the obtained numerical solutions for the shear force and moment along the dowel directly under the applied load. The analytical expressions for the shear and moment along the dowel, as shown in figures [3a] and [3b] respectively, were derived from equation [3], as presented by Friberg (1938). The 3D FE results for shear force and moment along the dowel were relatively close to the analytical solution. The solutions did not match exactly due to, among other things, the limitations involved in the assumptions behind the analytical solutions presented by Timoshenko (1925) and Frieberg (1938), among which: (1) the assumption of semi-infinite dowel length as opposed to actual dowels of finite size, (2) the assumption of semi-infinite elastic mass in which the dowel bar is embedded, as opposed to an actual layered pavement structure with finite thickness for each layer, and (3) the assumption of uniform modulus of dowel support along the dowel, as opposed to a varying K found from numerical results.

## Numerical Results

### Assumptions

The problem of modeling fatigue failure of PCC slabs under repetitive loading is computationally too expensive. To overcome this, a limit state approach was adopted, and it was assumed that the behavior of concrete in the vicinity of dowels under cyclic loading can be safely modeled with a single increasing overload. This was deemed acceptable as this study is oriented towards establishing a qualitative assessment of the initiation and propagation of damage in the vicinity of dowels leading ultimately to the failure of the PCC pavement structure. Following this line of though, a single pass of an overload was applied so as to cause full degradation of the concrete matrix around the central dowels, without changing the contact area of the applied load.

**General vs Local Concrete Deterioration**

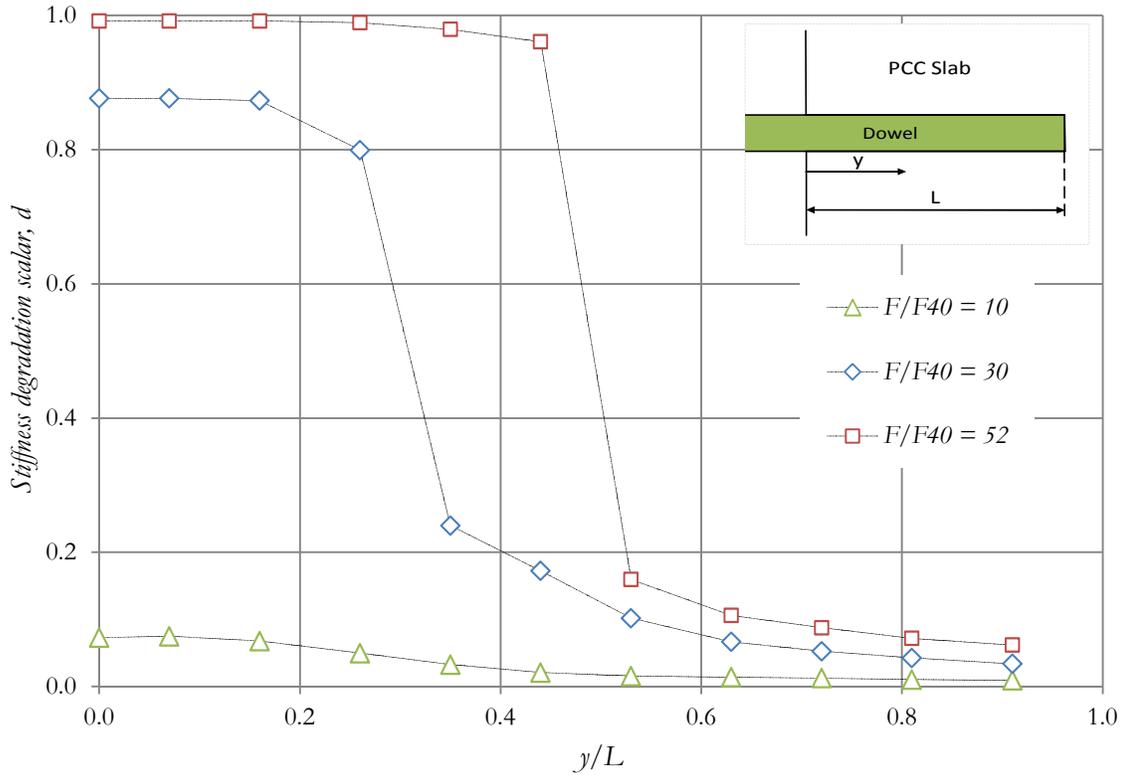

*Figure 4 – Stiffness degradation scalar d in the PCC around the dowel directly under the wheel load at three load levels $F/F_{40}$ of 56, 128, and 223.*

The purpose of this section is to describe the deterioration of PCC matrix around the dowel directly under the applied load. Figure [4] plots the distribution of the stiffness degradation parameter scalar parameter d along the dowel bar directly under the tire imprint at several load levels $F/F_{40}$. We recall that F is the wheel load applied in the form of pressure over a constant circular area. It can be observed that the highest level of degradation d occurs at the face of the joint (y/L=0). As the load level ($F/F_{40}$) increases, the degradation d was found to propagate towards the interior of the dowel bar. However, after a certain critical distance $(y/L)_c$, the amount of stiffness degradation d becomes practically negligible.

This leads to the idea that the deterioration of the PCC matrix is not uniform along the dowel. While the concrete at the face of the joint shows high levels of stiffness degradation, the concrete in the vicinity of the interior end of the dowel remains practically intact. Hence, the failure of concrete around dowels is a local rather than a general deterioration mechanism, as damage was found to originate at the face of the joint and propagate towards the interior of the dowel.

**Modulus of Dowel Support**

By definition, the modulus of dowel support K represents the stiffness of the elastic medium in which the dowel bar is embedded. Even though an accurate estimation of the modulus of dowel support is a key factor in the solutions presented by Friberg (1938), a sound theoretical method for estimating this parameter lacks. So far, the modulus of dowel support K has been determined through experimental testing, with values ranging from 81.5 to 409 N/mm$^{-3}$. The modulus of dowel support can be simply calculated by dividing the bearing stress in the concrete under the dowel by the corresponding vertical deflection.

In matching experimental results with analytical solution of Timoshenko, Mannava et al. (1999) noted that different values of K were required in order to match measured deflections along the dowel with Friberg's analytical solutions. This leads to the idea that a unique K-value along the dowel embedment length is incompatible with actual deflection measurements.

Figure [5] plots the stiffness degradation scalar in the concrete at four locations near the joint face corresponding to y/L equal to 0.0, 0.2, 0.4 and 0.6 as a function of the load level $F/F_{40}$. The results shown in this graph validate earlier findings that damage to the concrete matrix was confined to the face of the joint. In fact, as $F/F_{40}$ increases, damage at the joint face (y/L = 0.0) is always higher than towards the interior of the dowel. At y/L ≥ 0.6, damage in the concrete matrix is practically negligible; this remains true at higher load levels where the concrete at the joint face becomes totally degraded.

It should be noted that results from this numerical study did not match the assumption of uniform modulus of dowel support along the dowel that was adopted by Timoshenko (1925) and Friberg (1938) in deriving the analytical solutions. Values of k were not found to be uniform along the embedded parts of the dowel, even in the elastic range, and the practice of assigning a uniform modulus of dowel support k (calculated at the joint face) along the dowel should be revisited.

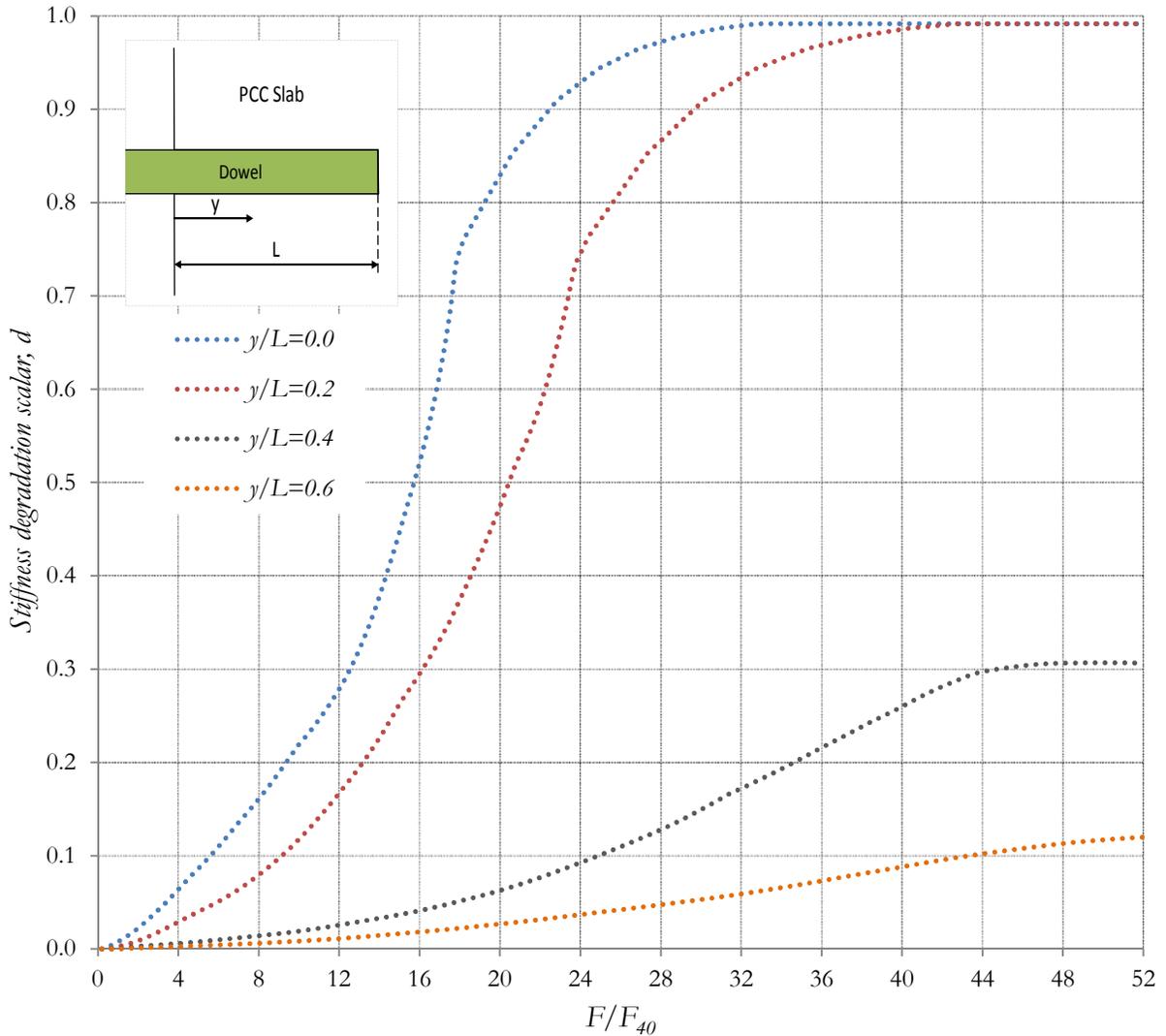

*Figure 5 – (a) Degradation of concrete matrix around central dowel towards the joint face (0≤y/L≤0.6).*

In addition, the modulus of dowel support K, at the joint face, was found to decrease significantly as $F/F_{40}$ increased. However, the same was not true at $y/L \geq 0.6$, where the modulus of dowel support maintained a relatively constant value with increasing $F/F_{40}$ values. This leads to the idea that the degradation of the modulus of dowel support K along the dowel is not uniform. Because the concrete material at the face of the joint suffers relatively greater damage than the corresponding concrete material towards the interior of the embedded dowel, the modulus of dowel support K at the face of the joint degrades more rapidly than the corresponding K along the interior of the dowel, particularly at $y/L \geq 0.6$.

**Load Transfer Capability & Dowel Group Action**

Figure [6a] shows the variation of maximum shear force (P) at the midpoint of each dowel as a function of the load level $F/F_{40}$. We recall that P also represents the amount of vertical load transferred by each dowel from one concrete slab to the next. Results are plotted for the dowel directly under the applied load and the two dowels directly next to the applied load. While the dowel under the applied load and the first dowel next to the wheel load carry vertical downward shear (P>0), the second dowel next to the applied load carries vertical upward shear, as shown with the negative P values. This may be explained by the fact that concrete slab curling resulting from the applied load may cause the edge dowels, which in this case is the second dowel next to the applied load, to transfer loads across slabs through upward shear action.

In figure [6b], the maximum shear force P at the midpoint of each dowel is normalized to the applied load F. As expected, as $F/F_{40}$ initially increases in the elastic range, the ratio P/F increases. This means that in the elastic range, as the applied load increases, the amount of shear force transferred by the dowel from one concrete slab to the other increases (relative to the applied wheel load). The dowel directly under the wheel load was found to transfer as much as approximately 15% of the applied load F.

However, once degradation of the concrete matrix surrounding the dowels under the applied load start, the capacity of the dowels to transfer vertical loads relative to the applied wheel load is reduced. For the dowel directly under the applied load, the normalized load transfer capability P/F reduces from approximately 15 % to 2% as the ratio $F/F_{40}$ increases to a value of 52.

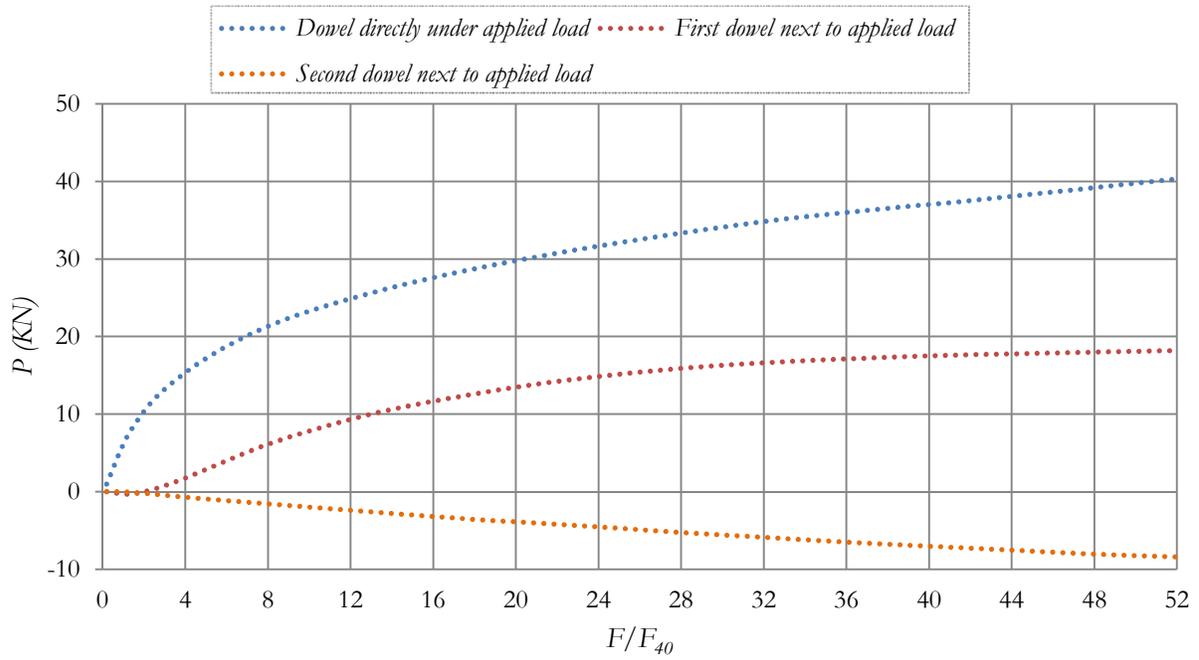

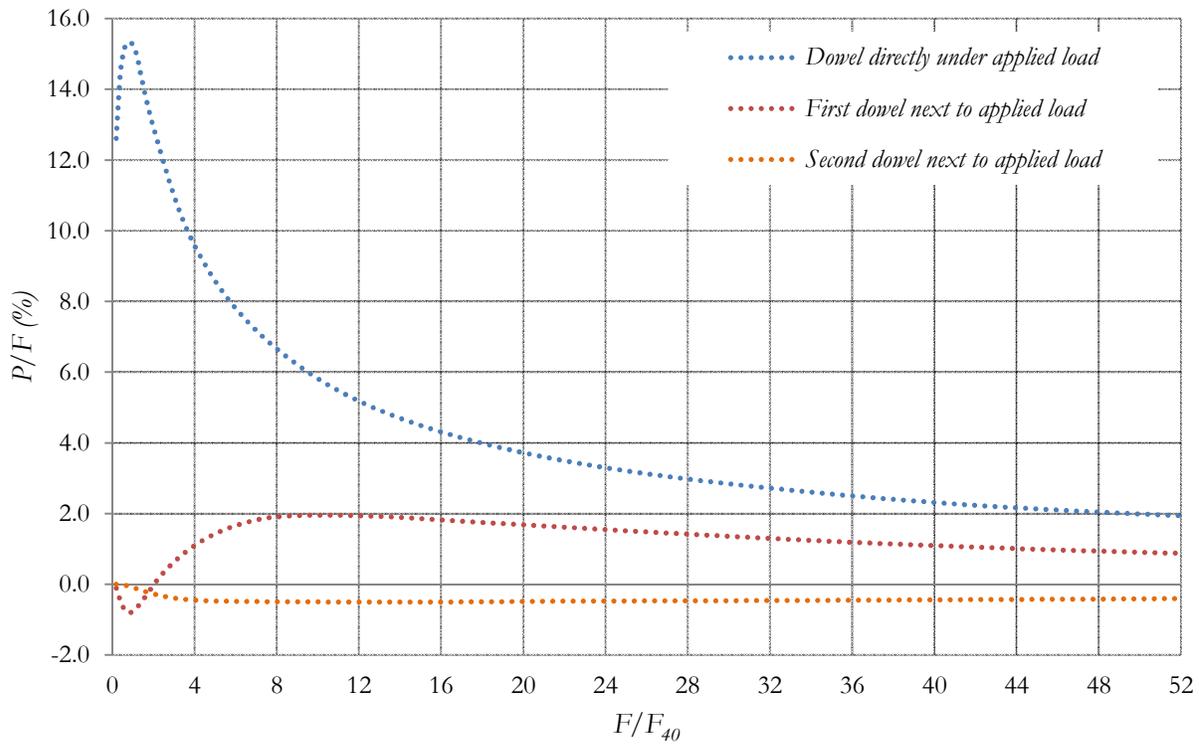

*Figure 6 - Evolution with load level of (a) maximum shear force P transferred by three dowels closest to the applied load, and (b) value of normalized shear force (P/F) in the three dowels.*

It should be also noted that once the material under the dowel directly under the applied load starts to degrade and the corresponding load transfer capacity P/F of the same dowel under the applied load drops, the applied load F is redistributed to the neighboring dowels. This can be visualized by the fact that as once the ratio P/F of the dowel directly under the applied load reaches its peak and starts to decrease, the corresponding ratio P/F for the first dowel next to the applied load starts to increase. In fact, at the same time where the central dowel directly under the wheel load is losing its load transfer capacity, the first dowel next to the wheel load becomes more engaged in load transfer. The normalized load transfer capability P/F of the first dowel next to the wheel load increases to a maximum value of approximately 2%. Then, once damage initiates along the concrete surrounding the first dowel next to the applied load, the load transfer capability P/F of this dowel starts to decrease from the maximum value of 2%, as shown in figure [6b].

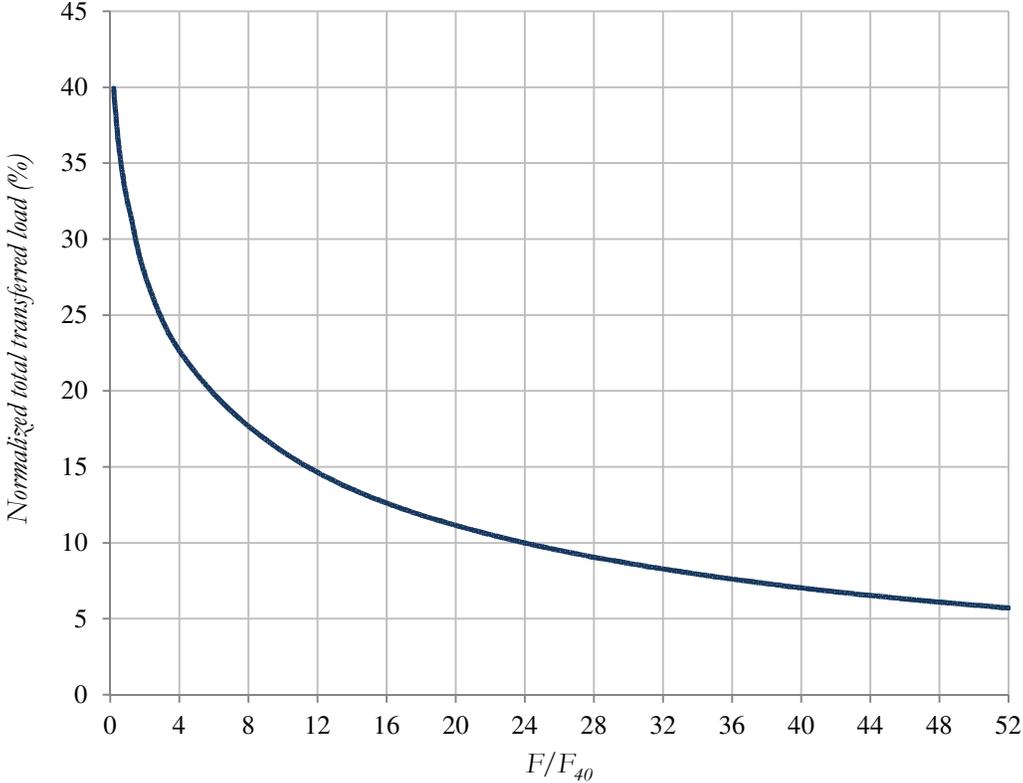

*Figure 7 - Variation of normalized total transferred load with load level. The normalized total transferred load is equal to the sum of maximum shear force transferred via all the transverse dowels divided by the applied load F.*

Figure [7] visualizes the evolution of load-carrying capacity of the dowel group with load level $F/F_{40}$. On the vertical axis, the normalized total transferred load represents the sum of maximum shear forces P carried by each dowel divided by the total applied load F, expressed as a percentage. It should be noted that the maximum value for the normalized total transferred load is theoretically 50 %. This is due to the fact that when the dowels are totally efficient, one-half of the applied load will be transferred from one slab to the other via dowel bars, and the remaining 50 % will be transferred through the base and subbase layers.

It can be observed that the normalized total transferred load exponentially decreases with load level $F/F_{40}$. The drop in the amount of total load transferred as the load level increases is mainly associated with the deterioration of the PCC matrix around the dowels. With increasing load level $F/F_{40}$ and the resulting degradation in the concrete matrix, the dowels partially lose their load-transfer capacity.

## Conclusions and Recommendations

In summary, this numerical study examined the deterioration of concrete matrix around dowel bars in typical rigid pavement structures. Using a nonlinear 3D FE model of a rigid pavement structure developed using ABAQUS, attempts were made to first describe the failure of dowel bars and investigate the distribution of the modulus of dowel support K along the dowel bar, and second, to describe the change in load transfer capacity of the transverse dowels as the surrounding PCC matrix degrades. The 3D FE model includes a damaged plasticity model for the concrete slabs, frictional interfaces, infinite boundary elements, and 3D beam model for dowel bars.

It can be concluded first that the degradation of concrete matrix around the dowel is a local rather than a general degradation process. Damage to the concrete surrounding the dowel, expressed in terms of stiffness degradation scalar d, was found to initiate at the face of the joint and propagate towards the interior of the dowel. The concrete matrix did not degrade uniformly along the embedded part of the dowel, and damage was confined to the face of the PCC joint.

Common practice is to uniformly reduce the modulus of dowel support along the dowel in order to model the deterioration of the dowel-concrete interface. However, in light of the results presented in this study, the modulus of dowel support K should be changed locally in modeling the deterioration of the concrete matrix around the dowels since degradation of the concrete was practically limited to the face of the joint.

In addition to that, results obtained show that as the concrete surrounding the central dowels under the applied load degraded, dowels directly under the applied load partly lose their load transfer capacity, while dowels farther away from the applied load become more engaged in load transfer. Overall, the dowels along the transverse joint lost a significant portion of their load-transfer capacity as the PCC matrix around them deteriorated.

Bhattacharjee, S.S., and Leger, P. (1993). "Seismic cracking and energy dissipation in concrete gravity dams." *Earthquake Engineering and Structural Dynamics*, Vol. 22, No. 11, pp. 991-1007.

Ghrib, F., and Tinawi, R., (1995). "An application of damage mechanics for seismic analysis of concrete gravity dams." *Earthquake Engineering and Structural Dynamics*, Vol. 24, No. 2, pp. 157-173.

Mannava, S.S., Bush, T.D., and Kukreti, A.R. (1999), "Load-deflection behavior of smooth dowels," Structural Journal, Vol. 96 (6), pp. 891 – 898.